\documentclass[aps,prb,showpacs,twocolumn,amsmath,amssymb]{revtex4}
\usepackage{graphicx}
\usepackage{bm}

\def\br{ \bm{r} }

\def\bk{ \bm{k} }

\def\bq{ \bm{q} }

\def\bgam{ \bm{\gamma} }

\def\tr{\,\mathrm{tr}\,}

\begin{document}

\title{Gap structure in noncentrosymmetric superconductors}

\author{K. V. Samokhin$^1$ and V. P. Mineev$^2$}

\affiliation{$^{1}$ Department of Physics, Brock University, St.Catharines,
Ontario L2S 3A1, Canada\\
$^{2}$ Commissariat \`a l'Energie Atomique, DSM/DRFMC/SPSMS, 38054
Grenoble, France}
\date{\today}

\begin{abstract}
Gap structure in noncentrosymmetric superconductors with
spin-orbit band splitting is studied using a microscopic model of
pairing mediated by phonons and/or spin fluctuations. The general
form of pairing interaction in the band representation is derived,
which includes both the intraband and interband pairing terms. In
the case of isotropic interaction (in particular, for a
BCS-contact interaction), the interband pairing terms vanish
identically at any magnitude of the band splitting. The effects of
pairing interaction anisotropy are analyzed in detail for a metal
of cubic symmetry with strong spin-orbit coupling. It is shown
that if phonons are dominant then the gaps in two bands are
isotropic, nodeless, and have in general different amplitudes.
Applications to the Li$_2$(Pd$_{1-x}$,Pt$_x$)$_3$B family of
noncentrosymmetric superconductors are discussed.
\end{abstract}


\maketitle

\section{Introduction}

Superconducting materials without inversion symmetry have recently
become a subject of considerable interest, both experimental and
theoretical. Starting from CePt$_3$Si (Ref. \onlinecite{Bauer04}),
the list of noncentrosymmetric superconductors has grown to
include UIr (Ref. \onlinecite{Akazawa04}), CeRhSi$_3$ (Ref.
\onlinecite{Kimura05}), CeIrSi$_3$ (Ref. \onlinecite{Sugitani06}),
Y$_2$C$_3$ (Ref. \onlinecite{Amano04}),
Li$_2$(Pd$_{1-x}$,Pt$_x$)$_3$B (Ref. \onlinecite{LiPt-PdB}),
KOs$_2$O$_6$ (Ref. \onlinecite{KOsO}), and other compounds. In
most cases, the fundamental questions about the gap symmetry and
the pairing mechanism remain unresolved.

The spin-orbit (SO) coupling of electrons with a
noncentrosymmetric crystal lattice lifts spin degeneracy of the
electron energy bands almost everywhere, which has important
consequences for superconductivity: In the limit of strong SO
coupling, the Cooper pairing between the electrons with opposite
momenta occurs only if they are from the same nondegenerate band.
This scenario is realized in CePt$_3$Si, where the SO band
splitting exceeds the critical temperature by orders of
magnitude.\cite{SZB04} The same is likely to be the case in other
materials, for instance in Li$_2$(Pd$_{1-x}$,Pt$_x$)$_3$B, see
Ref. \onlinecite{LP05}.

The pairing interaction between electrons is most naturally
introduced using the exact band
states,\cite{SZB04,GR01,SC04,Min04} which take into account all
the effects of the crystal lattice potential and the SO coupling,
see Sec. \ref{sec: Basics}. In the strong SO coupling limit, the
order parameter is represented by a set of complex functions, one
for each band, which makes the theory of noncentrosymmetric
superconductors similar to that of usual multiband
superconductors, see Ref. \onlinecite{SMW59}. An alternative
approach based on the representation of the pairing interaction in
terms of the pure spinor states unaffected by the SO coupling was
developed in Refs. \onlinecite{Edel89,FAKS04}.

In a phenomenological multiband pairing Hamiltonian, the relative
strength of pairing in different bands can be arbitrary. In this
article we go beyond the phenomenological description and study
the gap structure in noncentrosymmetric superconductors under some
fairly general assumptions about the microscopic mechanism of
pairing. Specifically, we consider the interaction mediated by
bosonic excitations (phonons and/or spin fluctuations). Starting
with a microscopic expression for a momentum and frequency
dependent pairing interaction, we derive the general form of the
pairing interaction in the band representation, which contains
both the intraband and interband pairing terms. The latter is
shown to vanish identically in the case of isotropic BCS-contact
interaction for any magnitude of the SO band splitting, see Sec.
\ref{sec: BCS}. In general, the interband pairing is absent only
in the limit of large band splitting, see Sec. \ref{sec:
Interaction}.

In Sec. \ref{sec: Cubic}, we present a detailed analysis of the
possible gap structures in noncentrosymmetric superconductors of
cubic symmetry, in a model which includes both the phonon and
spin-fluctuation mediated interactions. The Conclusion contains a
discussion of our results in the context of
Li$_2$(Pd$_{1-x}$,Pt$_x$)$_3$B experiments.

\section{Basic definitions}
\label{sec: Basics}

The Hamiltonian of noninteracting electrons in a
noncentrosymmetric crystal has the following form:
\begin{eqnarray}
\label{H_0}
    H_0=\sum\limits_{\bk}[\epsilon_0(\bk)\delta_{\alpha\beta}+
    \bgam(\bk)\bm{\sigma}_{\alpha\beta}]
    a^\dagger_{\bk\alpha}a_{\bk\beta}\nonumber\\
    =\sum_{\bk}\sum_{\lambda=\pm}\xi_\lambda(\bk)c^\dagger_{\bk\lambda}c_{\bk\lambda},
\end{eqnarray}
where $\alpha,\beta=\uparrow,\downarrow$ are spin indices,
$\bm{\sigma}$ are the Pauli matrices,
$\xi_\lambda(\bk)=\epsilon_0(\bk)+\lambda|\bgam(\bk)|$ are the
band dispersion functions, and the sum over $\bk$ is restricted to
the first Brillouin zone. In Eq. (\ref{H_0}) and everywhere below,
summation over repeated spin indices is implied, while summation
over the band indices is always shown explicitly. The SO coupling
of electrons with the crystal lattice is described by the
pseudovector $\bgam(\bk)$, which satisfies
$\bgam(-\bk)=-\bgam(\bk)$ and $(g\bgam)(g^{-1}\bk)=\bgam(\bk)$,
where $g$ is any operation from the point group $\mathbb{G}$ of
the crystal, see the examples below.

The Hamiltonian in the first line of Eq. (\ref{H_0}) is
diagonalized by the following transformation:
\begin{equation}
\label{band transform}
    a_{\bk\alpha}=\sum_{\lambda=\pm}u_{\alpha\lambda}(\bk)c_{\bk\lambda},
\end{equation}
with the coefficients
\begin{equation}
\label{Rashba_spinors}
    \begin{array}{l}
    \displaystyle u_{\uparrow\lambda}(\bk)=
    \sqrt{\frac{|\bgam|+\lambda\gamma_z}{2|\bgam|}},\\
    \displaystyle u_{\downarrow\lambda}(\bk)=\lambda
    \frac{\gamma_x+i\gamma_y}{\sqrt{2|\bgam|(|\bgam|+\lambda\gamma_z)}}
    \end{array}
\end{equation}
forming a unitary matrix $\hat u(\bk)$. The Fermi surfaces defined
by the equations $\xi_\pm(\bk)=0$ are split, except for the points
or lines where $\bgam(\bk)=0$. The band dispersion functions
$\xi_\lambda(\bk)$ are invariant with respect to all operations
from $\mathbb{G}$, and also even in $\bk$ due to time reversal
symmetry: the states $|\bk,\lambda\rangle$ and
$K|\bk,\lambda\rangle$ belong to $\bk$ and $-\bk$, respectively,
and have the same energy. Here $K=i\hat\sigma_2K_0$ is the time
reversal operation, and $K_0$ is the complex conjugation. One can
write $K|\bk,\lambda\rangle=t_\lambda(\bk)|-\bk,\lambda\rangle$,
where $t_\lambda(\bk)=-t_\lambda(-\bk)$ is a nontrivial phase
factor.\cite{GR01,SC04} For the eigenstates defined by expressions
(\ref{Rashba_spinors}) we obtain:
\begin{eqnarray}
\label{t lambda}
    t_\lambda(\bk)=\lambda
    \frac{\gamma_x(\bk)-i\gamma_y(\bk)}{\sqrt{\gamma_x^2(\bk)+\gamma_y^2(\bk)}}.
\end{eqnarray}

The momentum dependence of the SO coupling is determined by the
crystal symmetry. For the cubic group $\mathbb{G}=\mathbf{O}$,
which describes the point symmetry of
Li$_2$(Pd$_{1-x}$,Pt$_x$)$_3$B, the simplest form compatible with
the symmetry requirements is
\begin{equation}
\label{gamma_O}
    \bgam(\bk)=\gamma_0\bk,
\end{equation}
where $\gamma_0$ is a constant. For the point groups containing
improper elements, i.e. reflections and rotation-reflections,
expressions become more complicated. In the case of the full
tetrahedral group $\mathbb{G}=\mathbf{T}_d$, which is relevant for
Y$_2$C$_3$ and possibly KOs$_2$O$_6$, one has
\begin{equation}
\label{gamma_Td}
    \bgam(\bk)=\gamma_0[k_x(k_y^2-k_z^2)\hat x+k_y(k_z^2-k_x^2)\hat
    y+k_z(k_x^2-k_y^2)\hat z].
\end{equation}
This is also known as the Dresselhaus interaction,\cite{Dressel55}
and was originally proposed to describe the SO coupling in bulk
semiconductors of zinc-blende structure. For the tetragonal group
$\mathbb{G}=\mathbf{C}_{4v}$, which is relevant for CePt$_3$Si,
CeRhSi$_3$ and CeIrSi$_3$, the SO coupling is given by
\begin{equation}
\label{gamma C4v}
    \bgam(\bk)=\gamma_\perp(k_y\hat x-k_x\hat y)
    +\gamma_\parallel k_xk_yk_z(k_x^2-k_y^2)\hat z.
\end{equation}
In the purely two-dimensional case, setting $\gamma_\parallel=0$
one recovers the Rashba interaction,\cite{Rashba60} which is often
used to describe the effects of the absence of mirror symmetry in
semiconductor quantum wells.

Now let us take into account an attractive interaction between
electrons in the Cooper channel, using the basis of the exact
eigenstates of the noninteracting problem. The most general form
of the interaction Hamiltonian the band representation is
\begin{eqnarray}
\label{H int band gen}
    H_{int}=\frac{1}{2{\cal V}}\sum_{\bk\bk'\bq}\sum_{\lambda_{1,2,3,4}}
    V_{\lambda_1\lambda_2\lambda_3\lambda_4}(\bk,\bk';\bq)\nonumber\\
    \times c^\dagger_{\bk+\bq,\lambda_1}
    c^\dagger_{-\bk,\lambda_2}c_{-\bk',\lambda_3}
    c_{\bk'+\bq,\lambda_4}.
\end{eqnarray}
We assume that the $\bq$-dependence of the pairing interaction is
neglected (see the next section). The terms with
$\lambda_1=\lambda_2$ and $\lambda_3=\lambda_4$ describe intraband
pairing and the scattering of the Cooper pairs from one band to
the other, while the remaining terms describe pairing of electrons
from different bands. The above Hamiltonian can be considerably
simplified in the absence of the interband pairing, which is the
case if the SO splitting of the bands, $E_{SO}$, is large compared
with all energy scales associated with superconductivity. Since
the pairing interaction is effective only inside the shells of
width $\omega_c$ (the cutoff energy) in the vicinity of the Fermi
surfaces, one can set $\lambda_1=\lambda_2=\lambda$ and
$\lambda_3=\lambda_4=\lambda'$, and obtain:
\begin{eqnarray}
\label{H int reduced}
    H_{int}&=&\frac{1}{2{\cal V}}\sum\limits_{\bk\bk'\bq}\sum_{\lambda\lambda'}
    V_{\lambda\lambda'}(\bk,\bk')\nonumber\\
    &&\qquad\times c^\dagger_{\bk+\bq,\lambda}
    c^\dagger_{-\bk,\lambda}c_{-\bk',\lambda'}c_{\bk'+\bq,\lambda'},
\end{eqnarray}
where
\begin{equation}
\label{tilde V def}
    V_{\lambda\lambda'}(\bk,\bk')=t_\lambda(\bk)t^*_{\lambda'}(\bk')
    \tilde V_{\lambda\lambda'}(\bk,\bk').
\end{equation}
The pairing amplitudes $\tilde V_{\lambda\lambda'}$ are even in
both $\bk$ and $\bk'$ (due to the anticommutation of fermionic
operators) and also invariant under the point group operations:
$\tilde V_{\lambda\lambda'}(g^{-1}\bk,g^{-1}\bk')=\tilde
V_{\lambda\lambda'}(\bk,\bk')$.\cite{Sam04}

In the case of large SO band splitting, the order parameter has
only intraband components. It is uniform (in the absence of
external fields) and can be represented in the form
$\Delta_\lambda(\bk)=t_\lambda(\bk)\tilde\Delta_\lambda(\bk)$. The
gap functions $\tilde\Delta_\lambda$ transform according to one of
the even irreducible representations of the point group and
satisfy the following equations:
\begin{eqnarray}
\label{gap eq}
    \tilde\Delta_\lambda(\bk)=-T\sum_n\sum_{\lambda'}\int\frac{d^3\bk'}{(2\pi)^3}
    \tilde V_{\lambda\lambda'}(\bk,\bk')\nonumber\\
    \times\frac{\tilde\Delta_{\lambda'}(\bk')}{\omega_n^2+\xi_{\lambda'}^2(\bk')
    +|\tilde\Delta_{\lambda'}(\bk')|^2}.
\end{eqnarray}
The expression on the right-hand side converges due to the energy
cutoff at $\omega_c$.

\section{BCS model}
\label{sec: BCS}

Let us calculate the pairing amplitudes and the gap functions in a
simple BCS-like model, in which the attractive interaction is both
instantaneous in time and local in space:
\begin{eqnarray}
\label{H int BCS}
    H_{int}&=&-V\int d^3\br\,\psi_\uparrow^\dagger(\br)\psi_\downarrow^\dagger(\br)
    \psi_\downarrow(\br)\psi_\uparrow(\br)\nonumber\\
    &=&-\frac{V}{4}\int d^3\br\,(i\sigma_2)_{\alpha\beta}
    (i\sigma_2)^\dagger_{\gamma\delta}\nonumber\\
    &&\times\psi_\alpha^\dagger(\br)\psi_\beta^\dagger(\br)\psi_\gamma(\br)\psi_\delta(\br),
\end{eqnarray}
where $V>0$. Using the band representation of the field operators,
\begin{equation}
\label{psis}
    \psi_\alpha(\br)=\frac{1}{\sqrt{{\cal V}}}\sum_{\bk,\lambda}u_{\alpha\lambda}(\bk)
    e^{i\bk\br}c_{\bk\lambda},
\end{equation}
we obtain the pairing Hamiltonian in the form (\ref{H int band
gen}) with
\begin{eqnarray*}
    &&V_{\lambda_1\lambda_2\lambda_3\lambda_4}(\bk,\bk')=
    -\frac{V}{2}(i\sigma_2)_{\alpha\beta}(i\sigma_2)^\dagger_{\gamma\delta}\\
    &&\qquad\times u^*_{\alpha\lambda_1}(\bk)u^*_{\beta\lambda_2}(-\bk)
    u_{\gamma\lambda_3}(-\bk')u_{\delta\lambda_4}(\bk').
\end{eqnarray*}
Here we neglected the difference between
$u_{\alpha\lambda}(\pm\bk+\bq)$ and $u_{\alpha\lambda}(\pm\bk)$,
which is $\mathcal{O}(q/k_F)$. In conventional centrosymmetric
superconductors, we have $q/k_F\sim(\xi k_F)^{-1}\ll 1$ ($\xi$ is
the correlation length). In the noncentrosymmetric case, the above
estimate might not work and the $\bq$-dependence of the pairing
interaction might be more important, leading, for instance, to the
Lifshitz invariants in the free energy\cite{MS94,SZB04} and a
spatial modulation of the order parameter even in the absence of
external fields. We leave this issues to a separate publication.

Using the identities
\begin{equation}
\label{k minus k}
    u_{\alpha\lambda}(-\bk)=t^*_\lambda(\bk)(i\sigma_2)_{\alpha\beta}
    u^*_{\beta\lambda}(\bk),
\end{equation}
and also the unitarity of the matrix $\hat u(\bk)$, we obtain for
the pairing potential:
\begin{equation}
\label{V BCS}
    V_{\lambda_1\lambda_2\lambda_3\lambda_4}(\bk,\bk')=-\frac{V}{2}
    t_{\lambda_2}(\bk)t^*_{\lambda_3}(\bk')\delta_{\lambda_1\lambda_2}
    \delta_{\lambda_3\lambda_4}.
\end{equation}
Therefore, interband pairing is absent in the BCS model for any
strength of the SO coupling. Comparing this expression with Eq.
(\ref{tilde V def}), one can see that both the intraband pairing
and the pair scattering between the bands are characterized by the
same coupling constant: $\tilde
V_{\lambda\lambda'}(\bk,\bk')=-V/2$. The pairing symmetry is
isotropic, and it follows from Eqs. (\ref{gap eq}) that the gap
functions are the same in both bands:
$\tilde\Delta_+(\bk)=\tilde\Delta_-(\bk)=\eta$. This is not
surprising, since the local interaction (\ref{H int BCS}) cannot
lead to any $\bk$-dependence of the gaps.

The critical temperature is given by
$T_c=(2e^{\mathbb{C}}/\pi)\omega_c e^{-1/N_FV}$, where
$\mathbb{C}\simeq 0.577$ is Euler's constant, $N_F=(N_++N_-)/2$,
and $N_\lambda$ is the Fermi-level density of states in the
$\lambda$th band. Although this has the usual BCS form, the
superconductivity is non-BCS, because the order parameter resides
in two nondegenerate bands, with $T_c$ and $\eta$ independent of
the band splitting and the difference between $N_+$ and $N_-$. One
can show that both the critical temperature and the gap magnitude
are not affected by isotropic scalar impurities.\cite{MS07}

\section{Interaction mediated by bosonic excitations}
\label{sec: Interaction}

Now we investigate a more general model, in which the pairing is
assumed to be due to the exchange of some bosonic excitations. We
consider two types of excitations: scalar (phonons), which couple
to the electron density
$\rho(\br)=\psi^\dagger_\alpha(\br)\psi_\alpha(\br)$, and
pseudovector (spin fluctuations), which couple to the electron
spin density
$\bm{s}(\br)=\psi^\dagger_\alpha(\br)\bm{\sigma}_{\alpha\beta}\psi_\beta(\br)$.
Using the standard functional-integral representation of the
partition function of the system, we obtain the following term in
the fermionic action describing an effective two-particle
interaction between electrons:
\begin{eqnarray}
\label{S int xx}
    S_{int}&=&\frac{g_{ph}^2}{2}\int dx\,dx'
    \rho(x)D(x-x')\rho(x')\nonumber\\
    &&+\frac{g_{sf}^2}{2}\int dx\,dx'
    s_i(x){\cal D}_{ij}(x-x')s_j(x'),
\end{eqnarray}
where $x=(\br,\tau)$ is a shorthand notation for the coordinates
in real space and the Matsubara time, $\int dx(...)=\int
d\br\int_0^\beta d\tau(...)$, $g_{ph}$ and $g_{sf}$ are the
coupling constants of electrons with phonons and spin
fluctuations, while $D(x-x')$ and ${\cal D}_{ij}(x-x')$ are the
phonon and spin-fluctuation propagators respectively. The spin
fluctuations can be associated either with the localized spins, if
such are present in the system, or with the collective spin
excitations of the itinerant electrons
(paramagnons).\cite{paramagnons} In the latter case, ${\cal
D}_{ij}(x-x')$ can be expressed in terms of the electron dynamical
spin susceptibility $\chi_{ij}(\bq,\omega)$. In general, the
interaction (\ref{S int xx}) is nonlocal both in space and time.
The BCS-contact Hamiltonian (\ref{H int BCS}) is recovered when
the spin fluctuations are neglected and $g_{ph}^2D(\br,\tau)$ is
replaced by $-V\delta(\br)\delta(\tau)$.

In the momentum-frequency representation, Eq. (\ref{S int xx})
yields the following pairing action:
\begin{eqnarray}
\label{S int k}
    S_{int}=\frac{1}{2\Omega}\sum_{kk'q}\bigl[g_{ph}^2D(k-k')\delta_{\alpha\delta}
    \delta_{\beta\gamma}\nonumber\\
    +g_{sf}^2{\cal D}_{ij}(k-k')\sigma^i_{\alpha\delta}\sigma^j_{\beta\gamma}\bigr]
    \nonumber\\
    \times\bar a_\alpha(k+q)\bar
    a_\beta(-k)a_\gamma(-k')a_\delta(k'+q),
\end{eqnarray}
where $\Omega=\beta{\cal V}$ is the space-time volume, $\bar
a_\alpha(k)$ and $a_\alpha(k)$ are Grassmann fields,
$k=(\bk,\omega_n)$, $q=(\bq,\nu_m)$, and $\omega_n=(2n+1)\pi T$
and $\nu_m=2m\pi T$ are the fermionic and bosonic Matsubara
frequencies, respectively. We assume that the conditions of the
Migdal theorem are fulfilled, and also neglect the frequency
renormalization, which corresponds to the weak-coupling limit of
the Eliashberg theory. The theory developed below should work, at
least qualitatively, even for such materials as CePt$_3$Si, in
which strong electron correlations are responsible for a
heavy-fermion behavior and the above assumptions might be
inapplicable.

The phonon propagator is real and even in both frequency and
momentum: $D(k-k')=D(k'-k)$, and can therefore be written as
follows:
\begin{equation}
\label{D ug}
    D(k-k')=D^g(k,k')+D^u(k,k'),
\end{equation}
where the first term on the right-hand side,
$D^g(k,k')=[D(k-k')+D(k+k')]/2$, is even in both $k$ and $k'$,
while the second term, $D^u(k,k')=[D(k-k')-D(k+k')]/2$, is odd in
both $k$ and $k'$.

The spin-fluctuation propagator satisfies ${\cal
D}_{ij}(k-k')={\cal D}_{ji}(k'-k)$ and can be broken up into the
symmetric and antisymmetric in $ij$ parts. Representing the latter
in terms of a dual vector $\bm{R}$, we obtain
\begin{eqnarray}
\label{tilde D ug}
    {\cal D}_{ij}(k-k')={\cal D}^g_{ij}(k,k')+{\cal D}^u_{ij}(k,k')
    \nonumber\\
    +ie_{ijl}R_l(k-k'),
\end{eqnarray}
where the first (second) term on the right-hand side is an even
(odd) function of $k$ and $k'$, while $R_i(k-k')=-R_i(k'-k)$. The
antisymmetric component of the spin-fluctuation propagator is
associated with the Dzyaloshinskii-Moriya interaction.\cite{D-M}
It is absent in the centrosymmetric case, due to the additional
symmetry ${\cal D}_{ij}(k-k')={\cal D}_{ij}(k'-k)$.

After some straightforward algebra (see Appendix), the action
(\ref{S int k}) takes the following form:
\begin{eqnarray}
\label{S int gum}
    S_{int}&=&\frac{1}{2\Omega}\sum_{kk'q}V_{\alpha\beta\gamma\delta}(k,k')
    \nonumber\\
    &&\times\bar a_\alpha(k+q)\bar a_\beta(-k)a_\gamma(-k')a_\delta(k'+q),
\end{eqnarray}
where the pairing interaction is represented as a sum of the
$k$-even, $k$-odd, and mixed-parity terms: $V=V^g+V^u+V^m$. The
even contribution is
\begin{equation}
\label{V g}
    V^g_{\alpha\beta\gamma\delta}(k,k')=v_g(k,k')(i\sigma_2)_{\alpha\beta}
    (i\sigma_2)^\dagger_{\gamma\delta},
\end{equation}
where
\begin{equation}
\label{vg kk}
    v_g(k,k')=\frac{1}{2}\bigl[g_{ph}^2D^g(k,k')-g_{sf}^2\tr\hat{\cal D}^g(k,k')\bigr].
\end{equation}
The odd contribution is
\begin{equation}
\label{V u}
    V^u_{\alpha\beta\gamma\delta}(k,k')=
    v_{u,ij}(k,k')(i\sigma_i\sigma_2)_{\alpha\beta}(i\sigma_j\sigma_2)^\dagger_{\gamma\delta},
\end{equation}
where
\begin{eqnarray}
\label{vu kk}
    v_{u,ij}(k,k')=\frac{1}{2}\bigl[g_{ph}^2D^u(k,k')+g_{sf}^2
    \tr\hat{\cal D}^u(k,k')\bigr]\delta_{ij}\nonumber\\
    -g_{sf}^2{\cal D}^u_{ij}(k,k').
\end{eqnarray}
Finally, the mixed-parity contribution is
\begin{eqnarray}
\label{V m}
    V^m_{\alpha\beta\gamma\delta}(k,k')=v_{m,i}(k,k')
    (i\sigma_i\sigma_2)_{\alpha\beta}(i\sigma_2)^\dagger_{\gamma\delta}\nonumber\\
    +v_{m,i}(k',k)(i\sigma_2)_{\alpha\beta}
    (i\sigma_i\sigma_2)^\dagger_{\gamma\delta},
\end{eqnarray}
where
\begin{equation}
\label{vm kk}
    v_{m,i}(k,k')=\frac{g_{sf}^2}{2}[R_i(k-k')+R_i(k+k')].
\end{equation}
The first term on the right-hand side of Eq. (\ref{V m}) is odd in
$k$ and even in $k'$, while the second term is even in $k$ and odd
in $k'$.

We would like to note that expressions (\ref{V g}), (\ref{V u})
and (\ref{V m}) have completely general form in the sense that
they do not rely on our assumptions about boson-mediated
interactions and exhaust all possible spin structures of the
pairing amplitude. Under the point group operations $g$, the
coefficients $v_g$, $v_{u,ij}$, and $\bm{v}_m$ transform like a
scalar, a second-rank tensor, and a pseudovector, respectively,
and satisfy the invariance conditions
$v_g(g^{-1}\bk,\omega_n;g^{-1}\bk',\omega_{n'})=v_g(\bk,\omega_n;\bk',\omega_{n'})$,
\emph{etc}. By analogy with the theory of superconductivity in
centrosymmetric compounds, see e.g. Ref. \onlinecite{Book}, Eqs.
(\ref{V g}), (\ref{V u}) correspond to spin-singlet and
spin-triplet pairing channels respectively, while Eq. (\ref{V m})
describes singlet-triplet mixing. The possibility of
singlet-triplet mixing due to the Dzyaloshinskii-Moriya
interaction in the static case was pointed out in Ref.
\onlinecite{FAMS06}.

Next, we use Eqs. (\ref{band transform}) to transform the pairing
action into the band representation. Using identities (\ref{k
minus k}), we obtain the transformation rules for the pair
creation operators in the spin-singlet and spin-triplet channels:
\begin{eqnarray*}
    &&(i\sigma_2)_{\alpha\beta}\bar a_\alpha(k+q)\bar
    a_\beta(-k)\\
    &&\qquad=-\sum_{\lambda_{1,2}}t_{\lambda_2}(\bk)\delta_{\lambda_1\lambda_2}
    \bar c_{\lambda_1}(k+q)\bar c_{\lambda_2}(-k),\\
    &&(i\bm{\sigma}\sigma_2)_{\alpha\beta}\bar a_\alpha(k+q)\bar
    a_\beta(-k)\\
    &&\qquad=-\sum_{\lambda_{1,2}}t_{\lambda_2}(\bk)\bm{\tau}_{\lambda_1\lambda_2}(\bk)
    \bar c_{\lambda_1}(k+q)\bar c_{\lambda_2}(-k),
\end{eqnarray*}
where
\begin{equation}
    \hat\tau_i(\bk)=\hat u^\dagger(\bk)\hat\sigma_i\hat u(\bk).
\end{equation}
Inserting these in Eq. (\ref{S int gum}), we obtain:
\begin{eqnarray}
\label{S int band gen}
    &&S_{int}=\frac{1}{2\Omega}\sum_{kk'q}\sum_{\lambda_{1,2,3,4}}
    V_{\lambda_1\lambda_2\lambda_3\lambda_4}(k,k')\nonumber\\
    &&\qquad\times\bar c_{\lambda_1}(k+q)\bar c_{\lambda_2}(-k)
    c_{\lambda_3}(-k')c_{\lambda_4}(k'+q),\qquad
\end{eqnarray}
where
\begin{equation}
\label{V tilde V gen}
    V_{\lambda_1\lambda_2\lambda_3\lambda_4}(k,k')=
    t_{\lambda_2}(\bk)t^*_{\lambda_3}(\bk')
    \tilde V_{\lambda_1\lambda_2\lambda_3\lambda_4}(k,k'),
\end{equation}
and
\begin{eqnarray}
\label{tilde V gen}
    \tilde V_{\lambda_1\lambda_2\lambda_3\lambda_4}(k,k')=
    v_g(k,k')\delta_{\lambda_1\lambda_2}\delta_{\lambda_3\lambda_4}\nonumber\\
    +v_{u,ij}(k,k')\tau_{i,\lambda_1\lambda_2}(\bk)
        \tau_{j,\lambda_3\lambda_4}(\bk')\nonumber\\
    +v_{m,i}(k,k')\tau_{i,\lambda_1\lambda_2}(\bk)\delta_{\lambda_3\lambda_4}\nonumber\\
    +v_{m,i}(k',k)\delta_{\lambda_1\lambda_2}\tau_{i,\lambda_3\lambda_4}(\bk').
\end{eqnarray}
The pairing amplitudes satisfy the following symmetry properties:
\begin{eqnarray*}
    &&\tilde V_{\lambda_2\lambda_1\lambda_3\lambda_4}(-k,k')=
    \lambda_1\lambda_2\tilde
    V_{\lambda_1\lambda_2\lambda_3\lambda_4}(k,k'),\\
    &&\tilde V_{\lambda_1\lambda_2\lambda_4\lambda_3}(k,-k')=
    \lambda_3\lambda_4\tilde
    V_{\lambda_1\lambda_2\lambda_3\lambda_4}(k,k').
\end{eqnarray*}
To obtain these, we used the anticommutation of the Grassmann
fields in Eq. (\ref{S int band gen}) and also the expressions
(\ref{t lambda}) for the phase factors in Eq. (\ref{V tilde V
gen}).

It follows from Eq. (\ref{tilde V gen}) that, in general, all
possible channels are present in the pairing interaction,
including interband pairing. The latter is absent, for any
magnitude of the SO band splitting, if the odd harmonics of the
bosonic propagators are negligible, so that $v_{u,ij}(k,k')=0$ and
$v_{m,i}(k,k')=0$. This happens, in particular, for a fully
isotropic interaction, in which case
$v_g(k,k')=v_g(\omega_n,\omega_{n'})$.

We are particularly interested in the limit of large SO band
splitting, which is relevant for the majority of
noncentrosymmetric superconducting materials. In this limit, we
set $\lambda_1=\lambda_2=\lambda$ and
$\lambda_3=\lambda_4=\lambda'$ in Eq. (\ref{tilde V gen}) (the
case of arbitrary band splitting, with both intra- and interband
components of the order parameter present, will be considered in a
separate publication). Since
$\bm{\tau}_{\lambda\lambda}=\lambda\hat\bgam(\bk)$, the pairing
action becomes
\begin{eqnarray}
\label{S int band reduced}
    S_{int}=\frac{1}{2\Omega}\sum_{kk'q}\sum_{\lambda\lambda'}
    t_{\lambda}(\bk)t^*_{\lambda'}(\bk')\tilde V_{\lambda\lambda'}(k,k')\nonumber\\
    \times\bar c_{\lambda}(k+q)\bar c_{\lambda}(-k)
    c_{\lambda'}(-k')c_{\lambda'}(k'+q),
\end{eqnarray}
where
\begin{eqnarray}
\label{tilde V}
    \tilde V_{\lambda\lambda'}(k,k')=
    v_g(k,k')+\lambda\lambda'v_{u,ij}(k,k')\hat\gamma_i(\bk)\hat\gamma_j(\bk')\nonumber\\
    +\lambda\bm{v}_m(k,k')\hat\bgam(\bk)+\lambda'\bm{v}_m(k',k)\hat\bgam(\bk').
\end{eqnarray}
This expression, together with Eqs. (\ref{vg kk}), (\ref{vu kk})
and (\ref{vm kk}), relates the amplitudes of the intraband pairing
and the interband pair scattering to the bosonic excitation
spectra. Note that $\tilde V_{\lambda\lambda'}(k,k')$ is even in
both $k$ and $k'$. Treating the interaction (\ref{S int band
reduced}) in the mean-field approximation, see e.g. Ref.
\onlinecite{Book}, one introduces the order parameters
$\Delta_\lambda(k)=t_\lambda(\bk)\tilde\Delta_\lambda(\bk,\omega_n)$,
where, due to the symmetry of the pairing amplitudes,
$\tilde\Delta_\lambda(-\bk,-\omega_n)=\tilde\Delta_\lambda(\bk,\omega_n)$.

\subsection{Weak coupling model}
\label{sec: Weak coupling}

In order to make progress, we approximate the frequency dependence
of the pairing amplitudes by an anisotropic ``square-well''
model:\cite{AM82}
\begin{equation}
\label{square well}
    \tilde V_{\lambda\lambda'}(k,k')=\tilde V_{\lambda\lambda'}(\bk,\bk')
    \theta(\omega_c-|\omega_n|)\theta(\omega_c-|\omega_{n'}|)
\end{equation}
where $\theta(x)$ is the step function, $\omega_c$ is the
frequency cutoff, and $\tilde V_{\lambda\lambda'}(\bk,\bk')$
depend on the directions of $\bk$ and $\bk'$ near the
corresponding Fermi surfaces. The approximation (\ref{square
well}) has been used both for conventional phononic pairing
interaction, see Ref. \onlinecite{AM82}, and also for
spin-fluctuation mediated interaction, see Ref. \onlinecite{FA77}.
The ``square-well'' decomposition also holds for the gap
functions:
$\tilde\Delta_\lambda(\bk,\omega_n)=\tilde\Delta_\lambda(\bk)\theta(\omega_c-|\omega_n|)$,
so that the energy of quasiparticle excitations in the $\lambda$th
band is given by
\begin{equation}
\label{E k}
    E_\lambda(\bk)=\sqrt{\xi^2_\lambda(\bk)+|\tilde\Delta_\lambda(\bk)|^2}.
\end{equation}
The gap functions satisfy Eqs. (\ref{gap eq}), in which the
Matsubara sum is cut off at $\omega_c$.

The pairing amplitude given by the matrix Eq. (\ref{tilde V}) is
invariant under all operations from the crystal point group
$\mathbb{G}$, therefore $\tilde
V_{\lambda\lambda'}(g^{-1}\bk,g^{-1}\bk')=\tilde
V_{\lambda\lambda'}(\bk,\bk')$. Therefore, the momentum dependence
of each matrix element can be represented as a sum of the products
of the basis functions of irreducible representations of
$\mathbb{G}$. In general, the basis functions are different for
each matrix element. Neglecting this complication the pairing
amplitude can be factorized as follows:
\begin{equation}
\label{V expansion}
    \tilde V_{\lambda\lambda'}(\bk,\bk')=-\sum_a V^{a}_{\lambda\lambda'}
    \sum_{i=1}^{d_a}\phi_{a,i}(\bk)\phi_{a,i}^*(\bk'),
\end{equation}
where $a$ labels the irreducible representations (of
dimensionality $d_a$) of $\mathbb{G}$, which correspond to pairing
channels of different symmetry, with $\phi_{a,i}(\bk)$ being the
even basis functions.\cite{Book} The coupling constants
$V^{a}_{\lambda\lambda'}$ form a Hermitian matrix, which becomes
real symmetric if the basis functions are real. Keeping only the
irreducible representation $\Gamma$ which corresponds to the
maximum critical temperature, the gap functions take the form
\begin{equation}
\label{gaps}
    \tilde\Delta_\lambda(\bk)=\sum_{i=1}^{d_\Gamma}\eta_{\lambda,i}\phi_i(\bk),
\end{equation}
and $\eta_{\lambda,i}$ are the superconducting order parameter
components in the $\lambda$th band. The basis functions are
assumed to satisfy the following orthogonality conditions:
$\langle\phi^*_i(\bk)\phi_j(\bk)\rangle_\lambda=\delta_{ij}$,
where the angular brackets denote the averaging over the
$\lambda$th Fermi surface.

Linearizing the gap equations (\ref{gap eq}) we obtain the
following expression for the critical temperature:
\begin{equation}
\label{Tc}
    T_c=\frac{2e^{\mathbb{C}}}{\pi}\omega_ce^{-1/g},
\end{equation}
where
\begin{equation}
\label{g}
    g=\frac{g_{++}+g_{--}}{2}+\sqrt{\left(\frac{g_{++}-g_{--}}{2}\right)^2+g_{+-}g_{-+}}
\end{equation}
is the the effective coupling constant, and
\begin{equation}
\label{g def}
    g_{\lambda\lambda'}=V_{\lambda\lambda'}N_{\lambda'}.
\end{equation}
While the critical temperature is the same for all $d_\Gamma$
components of $\bm{\eta}_\lambda$, the gap structure in the
superconducting state below $T_c$, see Eq. (\ref{gaps}), is
determined by the nonlinear terms in the free energy, which
essentially depend on the symmetry of the dominant pairing
channel.

\section{Pairing symmetry in a cubic crystal}
\label{sec: Cubic}

In the case of isotropic pairing interaction, one can write
$v_g(k,k')=v_g(\omega_n,\omega_{n'})=-V_g\theta(\omega_c-|\omega_n|)\theta(\omega_c-|\omega_{n'}|)$
in the square-well approximation. In this way, one recovers the
BCS model of Sec. \ref{sec: BCS}, with $V=2V_g$ and the same
isotropic gaps in both bands.

To illustrate the effects of the interaction anisotropy on the gap
structure, let us consider the following example. In a cubic
crystal with $\mathbb{G}=\mathbf{O}$, the SO coupling can be
described by $\bgam(\bk)=\gamma_0\bk$. This model is applicable to
the Li$_2$(Pd$_{1-x}$,Pt$_x$)$_3$B family of noncentrosymmetric
compounds. The attractive interaction in these materials is likely
mediated by phonons,\cite{LP05,BZ05} therefore we neglect spin
fluctuations by setting $g_{sf}=0$ in expressions (\ref{vg kk}),
(\ref{vu kk}) and (\ref{vm kk}). Then,
$v_g(k,k')=(g_{ph}^2/2)D^g(k,k')$,
$v_{u,ij}(k,k')=(g_{ph}^2/2)D^u(k,k')\delta_{ij}$, and
$v_{m,i}(k,k')=0$. Using the square-well approximation, one has
\begin{eqnarray*}
    &&v_g(k,k')=v_g(\bk,\bk')\theta(\omega_c-|\omega_n|)\theta(\omega_c-|\omega_{n'}|),\\
    &&v_{u,ij}(k,k')=v_{u,ij}(\bk,\bk')\theta(\omega_c-|\omega_n|)\theta(\omega_c-|\omega_{n'}|),
\end{eqnarray*}
with the momentum dependence inherited from the phonon propagator.
Assuming a spherical Fermi surface and keeping only the $s$- and
$p$-harmonics in the phonon propagator, we obtain:
\begin{eqnarray}
\label{v ph}
    &&v_g(\bk,\bk')=-V_g,\nonumber\\
    &&v_{u,ij}(\bk,\bk')=-V_u(\hat{\bk}\hat{\bk}')\delta_{ij},\quad\\
    &&v_{m,i}(\bk,\bk')=0,\nonumber
\end{eqnarray}
where $V_g$ and $V_u$ are constants. Note that this interaction is
the same as the one considered phenomenologically by Edelstein in
Ref. \onlinecite{Edel89}. From Eq. (\ref{tilde V}) we obtain the
pairing amplitudes in the band representation as follows:
\begin{equation}
    \tilde V_{\lambda\lambda'}(\bk,\bk')=-V_g-\lambda\lambda'
    V_u(\hat{\bk}\hat{\bk}')^2.
\end{equation}
The components of the symmetric tensor $\hat k_i\hat k_j$
transform according to the representation $A_1+E+F_2$, where
$A_1$, $E$, and $F_2$ are respectively one-, two-, and
three-dimensional irreducible representations of the cubic group
$\mathbf{O}$ (the notations are the same as in Ref.
\onlinecite{LL3}). Therefore there are three pairing channels in
the expansion (\ref{V expansion}), with the following basis
functions and coupling constants:
\begin{eqnarray}
\label{V channels}
    &&V^{A_1}_{\lambda\lambda'}=V_g+\frac{1}{3}\lambda\lambda'V_u, \quad \phi_{A_1}(\bk)=1;\\
    &&V^{E}_{\lambda\lambda'}=\frac{2}{15}\lambda\lambda'V_u,\quad
    \bm{\phi}_{E}(\bk)\propto(\hat k_x^2+\omega\hat k_y^2+\omega^*\hat k_z^2,\nonumber\\
    &&\hspace*{4.35cm}\hat k_x^2+\omega^*\hat k_y^2+\omega\hat k_z^2);
    \qquad\nonumber\\
    &&V^{F_2}_{\lambda\lambda'}=\frac{2}{15}\lambda\lambda'V_u,
    \quad\bm{\phi}_{F_2}(\bk)\propto(\hat k_y\hat k_z,
    \hat k_z\hat k_x,\hat k_x\hat k_y),\nonumber
\end{eqnarray}
where $\omega=\exp(2\pi i/3)$.

Since phonons typically lead to a local attraction and cannot give
rise to a substantial $\bk$-dependence of the interaction, we
expect that the $A_1$ pairing channel dominates. Then the gap
functions in the two bands [Eqs. (\ref{gaps})] are isotropic:
$\tilde\Delta_\lambda(\bk)=\eta_\lambda$, and satisfy the
equations
\begin{equation}
\label{gap eq isotropic}
    \eta_\lambda=\sum_{\lambda'}g_{\lambda\lambda'}\,\pi
    T\sum_n\frac{\eta_{\lambda'}}{\sqrt{\omega_n^2+\eta^2_{\lambda'}}},
\end{equation}
where $g_{\lambda\lambda'}=V^{A_1}_{\lambda\lambda'}N_{\lambda'}$.
The critical temperature is given by Eq. (\ref{Tc}). The gap
magnitudes are not necessarily equal: For instance, in the
vicinity of $T_c$ we find the following expression for the gap
variation between the bands:
\begin{equation}
\label{gap variation}
    r\equiv\frac{\eta_+-\eta_-}{\eta_++\eta-}=
    \frac{g_{++}-g_{--}-2g_{-+}+\sqrt{\cal D}}{g_{++}-g_{--}+2g_{-+}+\sqrt{\cal D}},
\end{equation}
where ${\cal D}=\sqrt{(g_{++}-g_{--})^2+4g_{+-}g_{-+}}$. Assuming
that $N_+-N_-$ is small and that $V_g\gg V_u$, we have
\begin{equation}
\label{r phonons}
    r\simeq\frac{V_u}{6V_g}\frac{N_+-N_-}{N_F}.
\end{equation}
Thus the gaps are different only if an appreciable $p$-wave
harmonic is present in the phonon-mediated interaction \emph{and}
the SO coupling is sufficiently strong to create a considerable
difference between the densities of states in the two bands.

The coupling strengths being the same in both bands is not a
generic situation. In the spirit of the standard model of two-band
superconductivity,\cite{SMW59} it is possible that the coupling
constants corresponding to the intraband pairing channels and the
interband pair scattering are all different. To obtain this we
consider a generalization of the model (\ref{v ph}) which
includes, along with phonons, also a contribution from spin
fluctuations. In the absence of detailed information about the
phonon and spin-fluctuation spectra in real noncentrosymmetric
materials, in particular in Li$_2$(Pd$_{1-x}$,Pt$_x$)$_3$B, we use
the model which includes only the lowest angular harmonics
consistent with the symmetry requirements:
\begin{eqnarray}
    &&v_g(\bk,\bk')=-V_g,\nonumber\\
    &&v_{u,ij}(\bk,\bk')=-V_u(\hat{\bk}\hat{\bk}')\delta_{ij}-V_u'\hat k_i\hat k_j',\quad\\
    &&v_{m,i}(\bk,\bk')=-V_m\hat k_i.\nonumber
\end{eqnarray}
Here the coefficients $V_g$ and $V_u$ are, in general, different
from those in the model (\ref{v ph}). In the band representation,
the pairing amplitudes become
\begin{eqnarray}
    \tilde V_{\lambda\lambda'}(\bk,\bk')=-V_g-\lambda\lambda'
    \left[V_u(\hat{\bk}\hat{\bk}')^2+V'_u\right]\nonumber\\
    -(\lambda+\lambda')V_m.
\end{eqnarray}
There are three pairing channels, corresponding to the $A_1$, $E$
and $F_2$ representations, see Eqs. (\ref{V channels}). The
coupling constants in the $A_1$ channel now have the following
form:
\begin{equation}
\label{V A1 sf}
    V^{A_1}_{\lambda\lambda'}=V_g+\frac{1}{3}\lambda\lambda'V_u+
    \lambda\lambda'V'_u+(\lambda+\lambda')V_m.
\end{equation}
The gap functions are isotropic:
$\tilde\Delta_\lambda(\bk)=\eta_\lambda$, where the
$\eta_\lambda$s are found from equations (\ref{gap eq isotropic}).
The difference from the previous case is that now
$\eta_+\neq\eta_-$ even if the density of states variation between
the bands is negligible, i.e. $N_+=N_-=N_F$. Assuming that $V_m$
is smaller than the other constants (i.e. the singlet-triplet
mixing due to the Dzyaloshinskii-Moriya interaction is weak), we
obtain from Eq. (\ref{gap variation}) that
\begin{equation}
\label{r DM}
    r\simeq\frac{V_m}{V_g-V_u/3-V'_u}
\end{equation}
near the critical temperature.

Finally let us consider the case of $p$-wave interaction
dominating, which leads to an anisotropic pairing of the $F_2$
symmetry. This happens if $V_u$ is large enough, and the
degeneracy between the $F_2$ and $E$ channels is lifted, e.g., by
the Fermi surface anisotropy. The order parameter has the
following form:
\begin{equation}
\label{gaps F2}
    \tilde\Delta_\lambda(\bk)=\lambda\bigl(\eta_1\hat k_y\hat k_z+
    \eta_2\hat k_z\hat k_x+\eta_3\hat k_x\hat k_y\bigr).
\end{equation}
The symmetry of the gap, in particular the location of the nodes,
depends on the relation between the components of $\bm{\eta}$.
There are four stable states of a three-dimensional order
parameter in a cubic crystal:\cite{VG85} (i)
$\bm{\eta}=\eta_0(1,0,0)$, with two lines of nodes at $k_z=0$ and
$k_y=0$; (ii) $\bm{\eta}=\eta_0(1,i,0)$, with a line of nodes at
$k_z=0$, and also point nodes at $k_x=k_y=0$; (iii)
$\bm{\eta}=\eta_0(1,1,1)$, with two lines of nodes at the
intersection of the planes $\hat k_x+\hat k_y+\hat k_z=\pm 1$ with
the Fermi surface, and also point nodes at $k_x=k_y=0$,
$k_y=k_z=0$, and $k_z=k_x=0$; and (iv)
$\bm{\eta}=\eta_0(1,\omega,\omega^2)$, with point nodes at
$k_x=k_y=0$, $k_y=k_z=0$, $k_z=k_x=0$, and $k_x=k_y=k_z$. For the
first three states one would have $c_V(T)\propto T^2$ at low
temperatures,\cite{intersect} while for the last one
$c_V(T)\propto T^3$.

It is instructive to interpret our results using the spin
representation of the order parameter:
\begin{equation}
\label{Delta spin}
    \Delta_{\alpha\beta}(\bk)=\psi(\bk)(i\hat\sigma_2)_{\alpha\beta}+
    \bm{d}(\bk)(i\hat{\bm{\sigma}}\hat\sigma_2)_{\alpha\beta},
\end{equation}
where
\begin{equation}
\label{spin-singlet}
    \psi(\bk)=-\frac{\tilde\Delta_+(\bk)+\tilde\Delta_-(\bk)}{2}
\end{equation}
is the spin-singlet component, and
\begin{equation}
\label{spin-triplet}
    \bm{d}(\bk)=-\frac{\tilde\Delta_+(\bk)-\tilde\Delta_-(\bk)}{2}\hat\bgam(\bk)
\end{equation}
is the spin-triplet component.\cite{Min04,Sam07} The relative
strength of the triplet and singlet order parameters is controlled
by the difference between $\eta_+$ and $\eta_-$:
$|\bm{d}|/|\psi|=r$, see Eq. (\ref{gap variation}). In agreement
with Ref. \onlinecite{FAKS04}, only the component of $\bm{d}(\bk)$
which is parallel to $\hat\bgam(\bk)$ survives (is ``protected'')
in the limit of large SO band splitting. However, in the case of a
weakly anisotropic phonon-dominated interaction, it follows from
expression (\ref{r phonons}) that the triplet component is
negligibly small. In the opposite case, when the interaction is
strongest in the $p$-wave channel, one obtains from Eq. (\ref{V
channels}) that $\psi(\bk)=0$, i.e. the pairing is purely triplet.

\section{Conclusions and discussion}
\label{sec: Conclusion}

We have studied the pairing symmetry in noncentrosymmetric
superconductors with SO splitting of the electron bands. The
pairing interaction is derived using a microscopic model which
includes both phonons and spin fluctuations. The interband pairing
is shown to be absent for any strength of the SO coupling, if the
interaction anisotropy is negligible. We have analyzed possible
gap structures in the strong SO coupling limit with only intraband
pairing and interband pair scattering present, using a cubic
system as an example. If phonons are dominant, then the
superconducting gaps in both bands are isotropic and nodeless
(barring accidental zeros of the basis function of the unity
representation), but do not necessarily have the same magnitude.

Let us discuss the application of our results to the
noncentrosymmetric compounds Li$_2$(Pd$_{1-x}$,Pt$_x$)$_3$B, where
$x$ ranges from 0 to 1 (Ref. \onlinecite{LiPt-PdB}). The critical
temperature varies from 7-8 K for $x=0$ to 2.2-2.8 K for $x=1$.
The electronic band structure also exhibits considerable
variation: The SO band splitting in Li$_2$Pd$_3$B is as large as
30 meV, while in Li$_2$Pt$_3$B it reaches 200 meV (Ref.
\onlinecite{LP05}), which in both cases is much larger than $T_c$.
Due to the absence of strong correlation effects and magnetic
order, these materials provide a convenient testing ground for
theories of noncentrosymmetric superconductivity. Superconducting
pairing in Li$_2$Pd$_3$B is due to the exchange of phonons, and
the monotonic, almost linear, dependence of $T_c$ on the doping
level $x$ (Ref. \onlinecite{LiPt-PdB}) suggests that it remains
phononic for all $x$ from 0 to 1.\cite{LP05,BZ05}

Experimental data on the magnetic penetration depth,\cite{Yuan06}
the electronic specific heat,\cite{Takeya07}, and the NMR
characteristics,\cite{Nishi07} all seem to agree that
Li$_2$Pd$_3$B is a conventional BCS-like superconductor with no
gap nodes. In contrast, the gap structure in Li$_2$Pt$_3$B is
still a subject of intensive debates. While earlier experiments,
see Refs. \onlinecite{Yuan06,Takeya07,Nishi07}, suggested the
presence of lines of nodes in the gap, the recent $\mu$SR and
specific heat data\cite{Hafl07} have found no evidence of those.
Moreover, according to Ref. \onlinecite{Hafl07}, the whole
Li$_2$(Pd$_{1-x}$,Pt$_x$)$_3$B family of compounds are single-gap
isotropic superconductors. This conclusion is consistent with our
results, see Sec. \ref{sec: Cubic}. Indeed, assuming that the
pairing interaction in Li$_2$(Pd$_{1-x}$,Pt$_x$)$_3$B is phononic
and therefore only weakly anisotropic for all $x$, we obtain that
the $A_1$ channel always dominates, giving rise to nodeless
isotropic gaps of essentially equal magnitudes in both bands. In
order to create a noticeable difference between the gap
magnitudes, see Eq. (\ref{r phonons}), the interaction anisotropy
would have to be very strong: Since $(N_+-N_-)/N_F\sim
E_{SO}/\epsilon_F$ and varies from $0.03$ in Li$_2$Pd$_3$B to
$0.2$ in Li$_2$Pt$_3$B, the strength of the $p$-wave harmonic must
be at least an order of magnitude larger than that of the $s$-wave
harmonic, which is highly unlikely for a phonon-mediated
interaction.

\section*{ACKNOWLEDGEMENTS}

We thank B. Mitrovi\'c and S. Bose for useful discussions. The
financial support from the Natural Sciences and Engineering
Research Council of Canada (K.S.) is gratefully acknowledged.

\appendix

\section{Derivation of Eqs. (\ref{V g}-\ref{vm kk})}

Let us start from Eq. (\ref{S int k}), in which we substitute
expressions (\ref{D ug}) and (\ref{tilde D ug}):
\begin{eqnarray}
\label{A1}
    S_{int}=\frac{1}{2\Omega}\sum_{kk'q}\bigl\{g_{ph}^2[D^g(k,k')+D^u(k,k')]
    \delta_{\alpha\delta}\delta_{\beta\gamma}\nonumber\\
    +g_{sf}^2[{\cal D}^g_{ij}(k,k')+{\cal
    D}^u_{ij}(k,k')\nonumber\\
    +ie_{ijl}R_l(k-k')]\sigma^i_{\alpha\delta}\sigma^j_{\beta\gamma}\bigr\}
    \nonumber\\
    \times\bar a_\alpha(k+q)\bar
    a_\beta(-k)a_\gamma(-k')a_\delta(k'+q).
\end{eqnarray}
The $q$-dependence of the fermionic fields plays no role in the
algebraic transformations below, hence we use a shorter expression
on the right-hand side:
\begin{equation}
    S_{int}\to\frac{1}{2\Omega}\left(I^g+I^u+I^m\right),
\end{equation}
where
\begin{eqnarray*}
    I^g=\frac{1}{4}\sum_{kk'}\bigl[g_{ph}^2D^g(k,k')\delta_{\alpha\delta}
    \delta_{\beta\gamma}+g_{sf}^2{\cal D}^g_{ij}(k,k')
    \sigma^i_{\alpha\delta}\sigma^j_{\beta\gamma}\bigr]\\
    \times[\bar a_\alpha(k)\bar a_\beta(-k)-\bar a_\beta(k)\bar a_\alpha(-k)]\\
    \times[a_\gamma(-k')a_\delta(k')-a_\delta(-k')a_\gamma(k')],\\
    I^u=\frac{1}{4}\sum_{kk'}\bigl[g_{ph}^2D^u(k,k')\delta_{\alpha\delta}
    \delta_{\beta\gamma}+g_{sf}^2{\cal D}^u_{ij}(k,k')
    \sigma^i_{\alpha\delta}\sigma^j_{\beta\gamma}\bigr]\\
    \times[\bar a_\alpha(k)\bar a_\beta(-k)+\bar a_\beta(k)\bar a_\alpha(-k)]\\
    \times[a_\gamma(-k')a_\delta(k')+a_\delta(-k')a_\gamma(k')],\\
    I^m=\frac{1}{8}ie_{ijl}g_{sf}^2\sum_{kk'}[R_l(k-k')+R_l(k+k')]
    \sigma^i_{\alpha\delta}\sigma^j_{\beta\gamma}\\
    \times[\bar a_\alpha(k)\bar a_\beta(-k)+\bar a_\beta(k)\bar a_\alpha(-k)]\\
    \times[a_\gamma(-k')a_\delta(k')-a_\delta(-k')a_\gamma(k')]\\
    +\frac{1}{8}ie_{ijl}g_{sf}^2\sum_{kk'}[R_l(k-k')-R_l(k+k')]
    \sigma^i_{\alpha\delta}\sigma^j_{\beta\gamma}\\
    \times[\bar a_\alpha(k)\bar a_\beta(-k)-\bar a_\beta(k)\bar a_\alpha(-k)]\\
    \times[a_\gamma(-k')a_\delta(k')+a_\delta(-k')a_\gamma(k')].
\end{eqnarray*}
The even in $k$ combinations of the fermionic fields can be
represented as follows:
\begin{eqnarray}
    &&\bar a_\alpha(k)\bar a_\beta(-k)-\bar a_\beta(k)\bar a_\alpha(-k)\nonumber\\
    &&\quad\qquad=-(i\sigma_2)^\dagger_{\alpha\beta}(i\sigma_2)_{\mu\nu}
    \bar a_\mu(k)\bar a_\nu(-k),\nonumber\\
    &&a_\gamma(-k')a_\delta(k')-a_\delta(-k')a_\gamma(k')\nonumber\\
    &&\quad\qquad=-(i\sigma_2)_{\gamma\delta}(i\sigma_2)^\dagger_{\rho\sigma}
    a_\rho(-k')a_\sigma(k'),
\end{eqnarray}
while the odd combinations have the form
\begin{eqnarray}
    &&\bar a_\alpha(k)\bar a_\beta(-k)+\bar a_\beta(k)\bar a_\alpha(-k)\nonumber\\
    &&\quad\qquad=(i\sigma_i\sigma_2)^\dagger_{\alpha\beta}
    (i\sigma_i\sigma_2)_{\mu\nu}\bar a_\mu(k)\bar
    a_\nu(-k),\nonumber\\
    &&a_\gamma(-k')a_\delta(k')+a_\delta(-k')a_\gamma(k')\nonumber\\
    &&\quad\qquad=(i\sigma_i\sigma_2)_{\gamma\delta}(i\sigma_i\sigma_2)^\dagger_{\rho\sigma}
    a_\rho(-k')a_\sigma(k').
\end{eqnarray}
Using the matrix identities
\begin{eqnarray}
    &&\delta_{\alpha\delta}\delta_{\beta\gamma}
    (i\sigma_2)^\dagger_{\alpha\beta}(i\sigma_2)_{\gamma\delta}=2,\nonumber\\
    &&(\sigma_i)_{\alpha\delta}(\sigma_j)_{\beta\gamma}
    (i\sigma_2)^\dagger_{\alpha\beta}(i\sigma_2)_{\gamma\delta}=-2\delta_{ij},\nonumber\\
    &&\delta_{\alpha\delta}\delta_{\beta\gamma}(i\sigma_i\sigma_2)^\dagger_{\alpha\beta}
    (i\sigma_j\sigma_2)_{\gamma\delta}=2\delta_{ij},\nonumber\\
    &&(\sigma_i)_{\alpha\delta}(\sigma_j)_{\beta\gamma}(i\sigma_m\sigma_2)^\dagger_{\alpha\beta}
    (i\sigma_n\sigma_2)_{\gamma\delta}\\
    &&\qquad\qquad=2(\delta_{ij}\delta_{mn}-\delta_{im}\delta_{jn}-\delta_{in}\delta_{jm}),\nonumber\\
    &&(\sigma_i)_{\alpha\delta}(\sigma_j)_{\beta\gamma}(i\sigma_m\sigma_2)^\dagger_{\alpha\beta}
    (i\sigma_2)_{\gamma\delta}=2ie_{ijm},\nonumber\\
    &&(\sigma_i)_{\alpha\delta}(\sigma_j)_{\beta\gamma}(i\sigma_2)^\dagger_{\alpha\beta}
    (i\sigma_m\sigma_2)_{\gamma\delta}=-2ie_{ijm},\nonumber
\end{eqnarray}
we arrive at Eqs. (\ref{V g}-\ref{vm kk}).

\end{document}